\documentclass[aps,twocolumn,showpacs,prl,superscriptaddress]{revtex4-2}
\usepackage{mathtools}
\usepackage{bm}
\usepackage{dsfont,amsthm,amsbsy}
\usepackage{verbatim}
\usepackage{amssymb}
\usepackage{amsmath}
\usepackage{bbm}
\usepackage{graphicx}
\usepackage{epstopdf}
\usepackage{subfigure}
\usepackage{epsfig}
\usepackage{amsfonts}
\usepackage{mathrsfs}
\usepackage{sidecap}
\usepackage{lipsum}
\usepackage[toc,page,title,titletoc,header]{appendix}
\usepackage[colorlinks,linkcolor=blue,citecolor=blue,anchorcolor=blue,urlcolor=blue]{hyperref}
\usepackage{hyperref}
\usepackage{resizegather}
\usepackage{tikz}
\usepackage{float}
\usepackage{mathbbol}
\usepackage[normalem]{ulem}
\usepackage{cancel}
\usepackage{upgreek}

\newcommand{\bl}{\begin{aligned}}
\newcommand{\el}{\end{aligned}}

\def\be{\begin{equation}}
\def\ee{\end{equation}}

\def\bi{\begin{itemize}}
\def\ei{\end{itemize}}
\def\bn{\begin{enumerate}}
\def\en{\end{enumerate}}
\def\bea{\begin{eqnarray}}
\def\eea{\end{eqnarray}}
\def\no{\nonumber}
\def\ba{\begin{array}}
\def\ea{\end{array}}
\def\bd{\begin{displaymath}}
\def\ed{\end{displaymath}}

\begin{document}
\title{Scaling and Universality at Noise-Affected Non-Equilibrium Spin Correlation Functions}


\author{R. Jafari}
\email[]{raadmehr.jafari@gmail.com}
\affiliation{Department of Physics, Institute for Advanced Studies in Basic Sciences (IASBS), Zanjan 45137-66731, Iran}
\affiliation{School of Quantum Physics and Matter Science, Institute for Research in Fundamental Sciences (IPM), 19395-5531, Tehran, Iran}


%
\author{Alireza Akbari}
\email[]{alireza@bimsa.cn}
\affiliation{Beijing Institute of Mathematical Sciences and Applications (BIMSA), Huairou District, Beijing 101408, China}

\date{\today}

\begin{abstract}
We investigate scaling and universality in nonequilibrium spin correlation functions in the presence of uncorrelated noise. 
In the absence of noise, spin correlation functions exhibit a crossover from monotonic decay at fast sweep velocities to oscillatory behavior at slow sweeps. 
We show that, under a stochastically driven field, the critical sweep velocity at which the spin correlation functions undergo an abrupt change decreases with increasing noise strength and scales linearly with the square of the noise intensity. 
Remarkably, when the noise intensity and sweep velocity are comparable, the excitation probability becomes locked to $p_k = 1/2$ over a finite momentum window, signaling the emergence of noise-induced maximally mixed modes. 
This gives rise to a highly oscillatory region in the dynamical phase diagram, whose threshold sweep velocity increases with noise and likewise exhibits quadratic scaling with the noise strength. 
Finally, we identify a universal scaling function under which all boundary sweep-velocity curves collapse onto a single universal curve.
\end{abstract}

\maketitle
{\em Introduction-} 
The concepts of scaling and universality are fundamental to our understanding of equilibrium critical phenomena and are naturally captured within the framework of the renormalization group \cite{Cardy1996,Sachdev1999}.
This concept has been extended to nonequilibrium classical systems, leading to the identification of novel dynamical universality classes, including surface growth, coarsening, and reaction--diffusion processes \cite{Tauber2017}.
Recent advances in experiments on quantum many-body systems call for an extension of this concept to encompass universal phenomena in nonequilibrium quantum systems. Such universal behavior has been observed in driven open quantum systems, including experiments on nonequilibrium Bose--Einstein condensation of polaritons \cite{kasprzak2006}, dynamical phase diagrams of condensates trapped in optical cavities \cite{Kessler2019,Klinder2015}, and dissipative phase transitions in cavity QED circuits \cite{Fitzpatrick2017}. In addition, experiments with ultracold atoms and ions have revealed other types of dynamical transitions \cite{Smale2019,Zhang2017}, as well as new forms of dynamical scaling \cite{Nicklas2015,Prufer2018,Erne2018}.
Moreover, new classes of universal phenomena in nonequilibrium systems, known as dynamical quantum phase transitions (DQPTs), have been experimentally observed in several recent studies \cite{Karrasch2013,Karrasch2017,Kennes2018,Flaschner2018,Jurcevic2017,Bhattacharya2017,Martinez2016,Guo2019,Wang2019,Nie2020,Tian2020}, thereby confirming earlier theoretical predictions \cite{Heyl2013,Heyl2018,Heyl2015,Abeling2016,Khatun2019,Uhrich2020,Vanhala2023,Mondal2023,Cao2020,Sedlmayr2023,Sedlmayr2022,Sedlmayr2020,Zeng2023,Stumper2022,Yu2021,Vijayan2023,Bhattacharjee2023,Leela2024,Puskarov2016,Sacramento2024,Haldar2020,Zamani2020,Kosior2018,Kosior2018b,Naji2022b,Somendra2024,Jing2024,Ding2020,Cao2024,Cao2024b,Mondal2022,Bandyopadhyay2021,Peotta2021,Nicola2022,Nicola2021,khan2023,Ye2024}.

Recently, it has been shown that in nonequilibrium spin systems the asymptotic behavior of spin correlation functions at large spatial separations undergoes an abrupt transition, at a critical sweep velocity, from monotonic to oscillatory decay \cite{Cherng2006}. Remarkably, this transition occurs at the same sweep velocity that corresponds to the maximal phase space of modes experiencing the strongest decoherence. In the vicinity of this critical sweep velocity, the parameters characterizing the correlation functions exhibit pronounced nonanalytic behavior.
Although spin correlation functions have been extensively studied in various spin systems \cite{Cherng2006,LSM1961,Barouch1970,Barouch1971,Konar2025,Roy2019,Chanda2016,Lakkaraju2024},
it remains an open question how their behavior is affected by noise, despite the ubiquity of stochastic driving in realistic quantum platforms.
Noise is ubiquitous and unavoidable in physical systems, arising naturally as an effective description of the dynamics of systems interacting with nonequilibrium environments or subjected to externally driven fields.

In this Letter, we demonstrate that spin correlation functions continue to exhibit an abrupt change even in the presence of uncorrelated noise added to the driving field. We show that the critical sweep velocity at which the spin correlation functions become singular decreases with increasing noise strength and scales linearly with the square of the noise intensity. Moreover, noise gives rise to an additional region in the dynamical phase diagram where the spin correlation functions display highly oscillatory behavior, revealing a counterintuitive role of noise in enhancing dynamical structure rather than merely suppressing coherence.
\\

%
\begin{figure*}[t]
\centerline{
\includegraphics[width=0.24\linewidth]{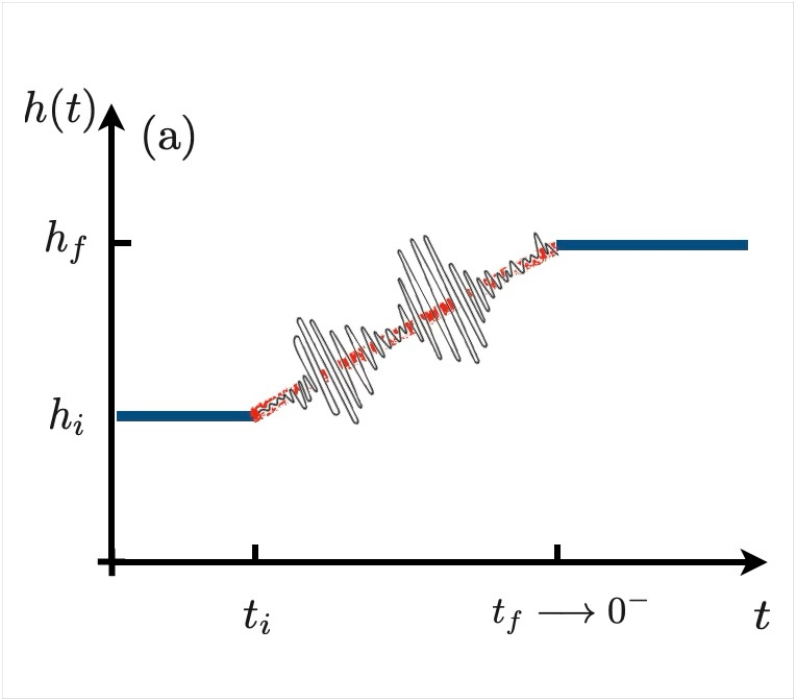}
\includegraphics[width=0.25\linewidth]{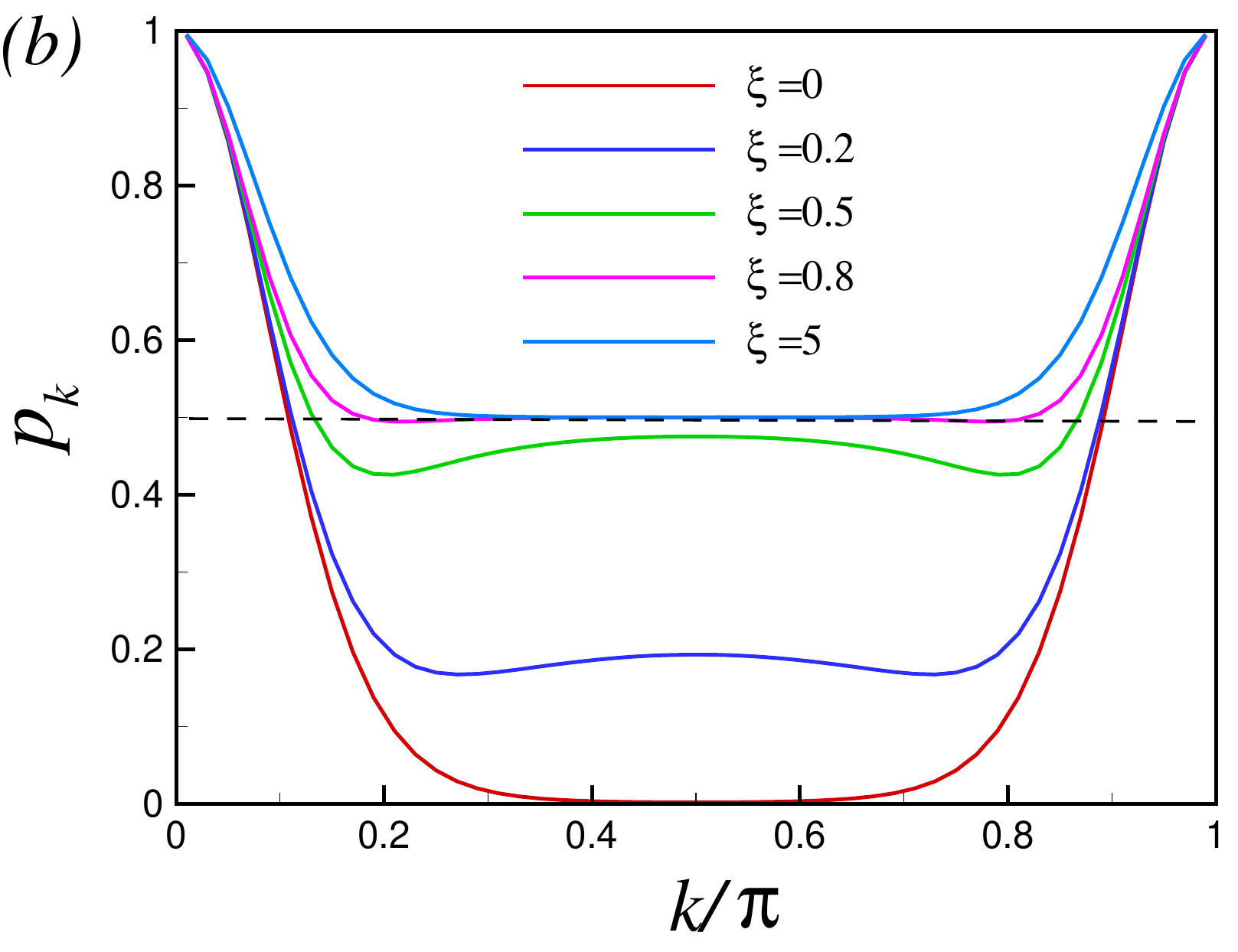}
\includegraphics[width=0.25\linewidth]{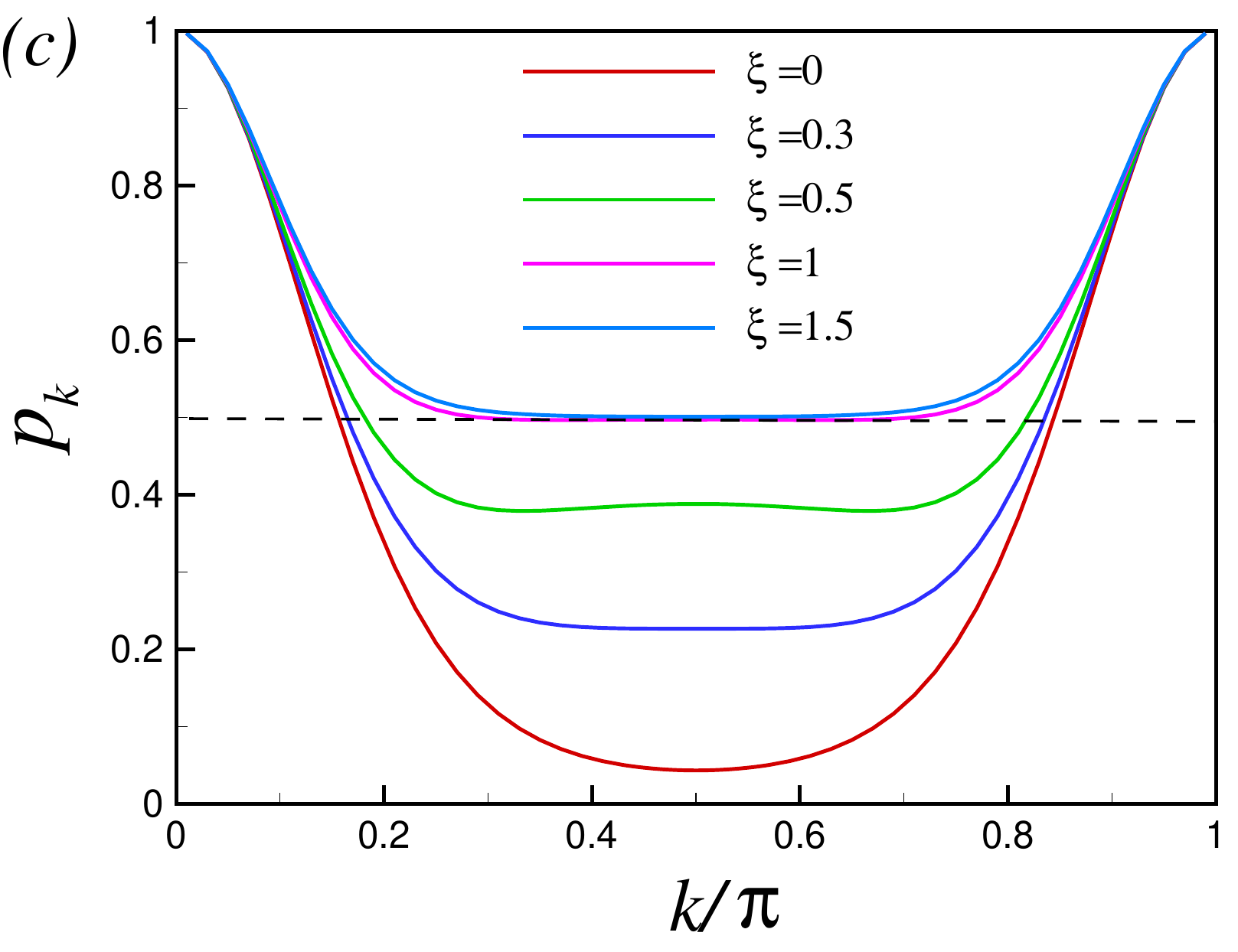}
\includegraphics[width=0.25\linewidth]{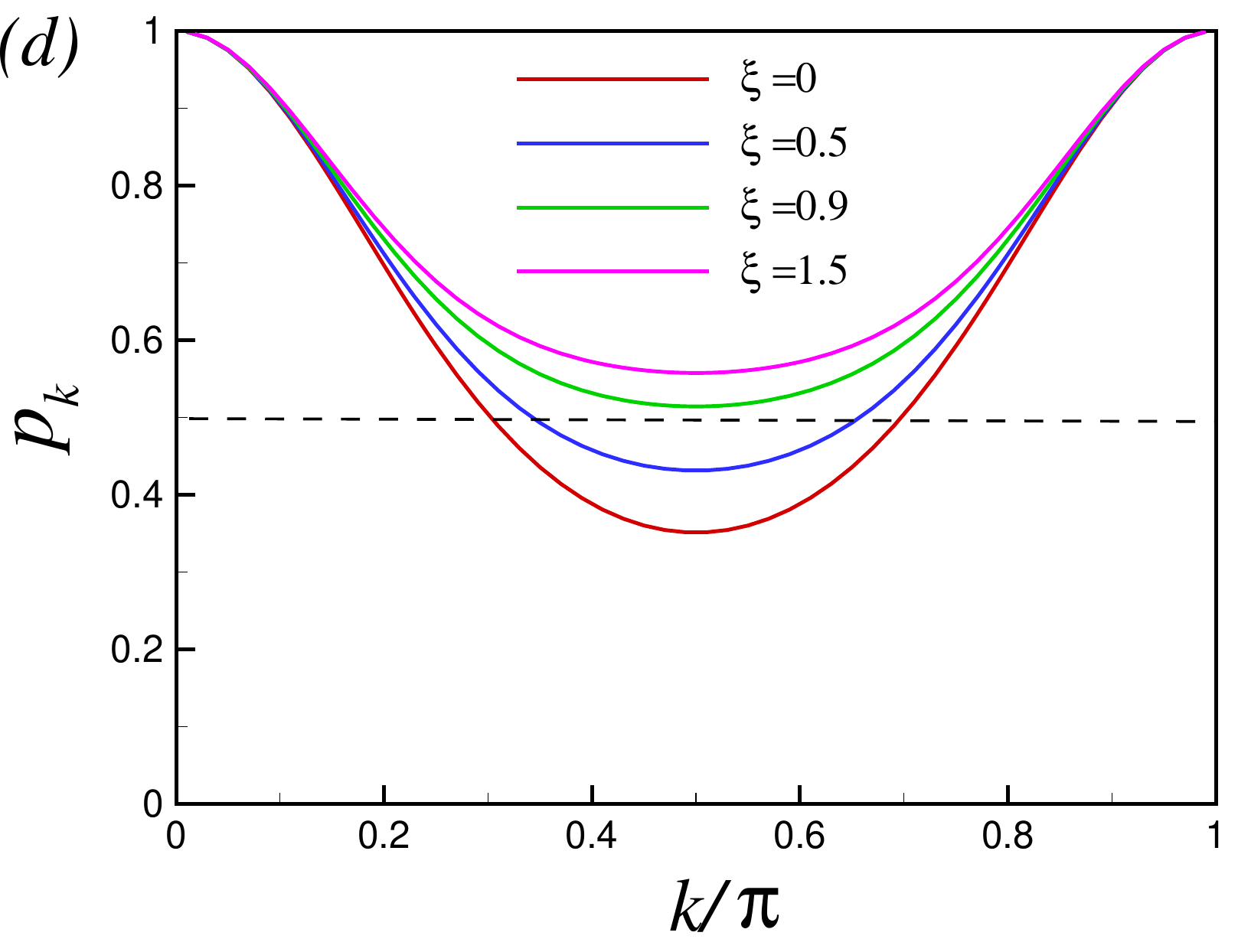}}
\vspace{-0.25cm}
\caption{
(a) Schematic illustration of a linear ramp quench (red), with superposed noise fluctuations (gray). Here, $h(t)$ denotes the magnetic field, with $h_i$ and $h_f$ its initial and final values, and $t_i$ and $t_f = 0$ the corresponding times.
Panels (b)–(d) show the probabilities $p_k$ of finding a fermionic mode with momentum $k$ in the upper band after a ramp across two quantum critical points at $h_c = \pm 1$ ($h_i = -100$, $h_f = 100$), for different noise amplitudes $\xi$ and sweep velocities: (b) $v = 0.5$, (c) $v = 1$, and (d) $v = 3$.}
\label{fig1}
\end{figure*}
%

{\em Model, ramp quench, and spin--spin correlation functions.-}
We consider the $XY$ model driven by a linearly ramped, noiseless transverse magnetic field $h_0(t)=vt$, with $t \in [t_i,t_f)$. The Hamiltonian reads
\begin{equation}
{\cal H}_0(t)
=
-\frac{J}{2} \sum_{j=1}^{N} \Bigg[
\frac{1+\gamma}{2}\,\sigma_{j}^x \sigma_{j+1}^x
+
\frac{1-\gamma}{2}\,\sigma_{j}^y \sigma_{j+1}^y
-
h_0(t)\,\sigma_{j}^z
\Bigg],
\end{equation}
with periodic boundary conditions. Throughout this work, we set $J=1$ as the energy scale and fix the system size to $N=500$.
The Hamiltonian ${\cal H}_0(t)$ can be mapped onto a model of spinless fermions with operators $c_j$, $c_{j}^{+}=\tau_{n-1}\sigma^{+}_{j}$, 
with $\tau_{j-1}=\prod_{p=1}^{j-1}\sigma_{p}^{z}$, and $\sigma^{\pm}_{j}=(\sigma^{x}_j\pm i\sigma^{y}_n)$ \cite{LSM1961}. 
Performing a Fourier transformation, $c_j = (1/\sqrt{N}) \sum_k \mbox{e}^{ikn}c_k$, and introducing the Nambu spinors $C_k^{\dagger} = (c_k^{\dagger} \ c_{-k})$, the Hamiltonian ${\cal H}_0(t)$ expressed as a sum over decoupled mode Hamiltonians 
${\cal H}_{0,k}(t)$,
%
\bea
\label{eq:Nambu}
{\cal H}_0(t) = \sum_{k} C^{\dagger}_k {\cal H}_{0,k}(t) C_k;
\quad
{\cal H}_{0,k}(t)=
\left(
\begin{array}{cc}
h_{z} & h_{x} \\
h_{x} & -h_{z}\\
\end{array}
\right),
\eea
%
where $h_{z}(k,t) = h_0(t)-\cos(k)$, $h_{x}(t) = \gamma\sin(k)$ with $k= (2m-1)\pi/N$ for $m=1, 2, \ldots, N/2$, and 
consequently the density matrix $\rho_0(t)$ has a direct product structure i.e., $\rho_{0}(t) = \otimes_k \,\rho_{0,k}(t)$ where
each $\rho_{0,k}(t)$ obeys the von-Neumann equation $\dot{\rho}_{0,k}(t)=-i[H_{0,k}(t),\rho_{0,k}(t)]$.
The instantaneous eigenstates and eigenvalues of ${\cal H}_{0,k}(t)$ are given by
%
\bea
\label{eqsm5}
|\phi^{-}_{k}(t)\rangle&=&\cos(\frac{\theta_k(t)}{2})|\alpha\rangle-i\sin(\frac{\theta_k(t)}{2})|\beta\rangle,\\
\no
|\phi^{+}_{k}(t)\rangle&=&-i\sin(\frac{\theta_k(t)}{2})|\alpha\rangle+\cos(\frac{\theta_k(t)}{2})|\beta\rangle,
\eea
%
and $\varepsilon_k^{\pm}(t)
=
\pm\varepsilon_k(t)=\pm\sqrt{h^{2}_{z}(k,t)+h_x(k)^2},$
respectively.
%
Here, $|\alpha\rangle = (1\, 0)^T$, $|\beta\rangle = (0\, 1)^T$, and $\theta_k(t)=\arctan[\Delta_k/h_k(t)]$. 
When the field is time-independent, $h_0(t) = h$, it is straightforward to show that, for anisotropy $\gamma\neq 0$, the gap of bands vanishes at 
$h_c=-1$ and $h_c=1$, with ordering wave vectors $k=\pi$ and $k=0$, respectively. These two lines correspond to the quantum phase transitions 
from a quantum paramagnetic phase to a ferromagnetically ordered phase with the associated exponents being the same as the transverse Ising model \cite{Barouch1970,Barouch1971}.

We focus on physically relevant observables, namely the spin correlation functions $\langle\sigma^{x(y)}_{\ell}\sigma^{x(y)}_{\ell+r}\rangle$ and the string correlator $\langle\tau_{\ell}\tau_{\ell+r}\rangle$, rather than on the full many-body density matrix.
The correlation functions can be written in the form~\cite{Barouch1971,Cherng2006,Jafari2025a,Naji2025,Jafari2026a}
%
\begin{equation}
\bl
\langle\sigma^{x}_{\ell}\sigma^{x}_{\ell+r}\rangle 
=& \langle B_{\ell}A_{\ell+1}B_{\ell+1} \cdots A_{\ell+r-1}B_{\ell+r-1}A_{\ell+r}\rangle,
\\
\langle\sigma^{y}_{\ell}\sigma^{y}_{\ell+r}\rangle 
=& \langle A_{\ell}A_{\ell+1}B_{\ell+1} \cdots A_{\ell+r-1}B_{\ell+r-1}B_{\ell+r}\rangle,
\el
\end{equation}
%
where $A_m = c_m^{\dagger} + c_m$, $B_m = c_m^{\dagger} - c_m$, and expectation values are defined as
$\langle \mathcal{O} \rangle = \mathrm{Tr}[\rho(t)\mathcal{O}]$.
It can be shown that the correlation functions can be expressed as a Pfaffian of a $2r \times 2r$ skew-symmetric matrix $C(r,t)$ \cite{Barouch1971},
%
\begin{equation}
\label{cormatrix}
C(r,t)=pf \left(
    \begin{array}{cc}
      S & G \\
      -G^T & Q \\
    \end{array}
  \right),
\end{equation}
%
where the elements of $S$ are functions of $\langle A_i A_j\rangle$, those of $Q$ are functions of $\langle B_i B_j\rangle$, and those of $G$ are functions of $\langle A_i B_j\rangle$ \cite{Jafari2026a}.
If the system is prepared in its ground state at $t_i$, then after a lengthy calculation one can show that \cite{Jafari2025a,Naji2025,Jafari2026a}
%
\begin{equation}
\bl
\label{eq:ABfunc}
&
\langle A_{\ell} A_{\ell+r} \rangle 
= \frac{1}{N} \sum_{k} e^{ikr} 
\mbox{Re}[\rho^{(d)}_{k,12}(t)] + \delta_{r,0}, 
\\
&
\langle B_{\ell} B_{\ell+r} \rangle
= \frac{1}{N} \sum_{k} e^{ikr} 
\mbox{Re}[\rho^{(d)}_{k,12}(t)]
- \delta_{r,0}, 
\\
&
\langle A_{\ell} B_{\ell+r} \rangle 
= - \langle B_{\ell+r} A_{\ell} \rangle 
\\
&
= \frac{1}{N} \sum_{k} e^{ikr} e^{2i\theta(t)}\Big[2 i \mbox{Im}[\rho^{(d)}_{k,12}(t)]+ \rho^{(d)}_{k,11}(t)-\rho^{(d)}_{k,22}(t) \Big],
\el
\end{equation} 
%
in which $\rho^{(d)}_{k,11}(t)=\langle\phi^{-}_k|\rho_k(t)|\phi^{-}_k\rangle$ and
$\rho^{(d)}_{k,22}(t)=\langle\phi^{+}_k|\rho_k(t)|\phi^{+}_k\rangle$~
\cite{Jafari2025a,Naji2025,Jafari2026a}.
It is straightforward to show that, for a linear ramp $h_0(t)=vt$ from an initial time $t_i \to -\infty$ ($h_i \ll h_c=-1$) to a final time $t_f \to +\infty$ ($h_f \gg h_c=1$), the summation $\sum_k e^{ikr}\rho^{(d)}_{k,12}(t)$ vanishes. Consequently, Eq.~(\ref{eq:ABfunc}) reduces to \cite{Cherng2006,Dziarmaga2022,Jafari2025a,Nag2011,Naji2025,Jafari2026a}

\begin{equation}
\label{eq:ABfunc2}
\bl
\langle A_{\ell} A_{\ell+r} \rangle 
=- \langle B_{l} B_{l+r} \rangle =
&\;
\delta_{r,0}, 
\\
\langle A_{\ell} B_{\ell+r} \rangle = - \langle B_{\ell+r} A_{\ell} \rangle
%
=
&
\frac{1}{N} \sum_{k} e^{ikr} (1-2p_{k}),
  \el  
\end{equation} 
%
where $p_{k}(t)=\rho^{(d)}_{k,22}=1-\rho^{(d)}_{k,11}$ is the excitation probability of the system at the end of ramp field $h_f$
which is given by Landau-Zener transition probability 
$p_k=\exp[-\pi(\gamma\sin(k))^2/v]$
for the noiseless ramp~\cite{Landau,Zener,Cherng2006,Zamani2024}. 
In such a case, the Pfaffian of correlation matrix, Eq. (\ref{cormatrix}), reduces to the $r\times r$ Toeplitz determinant of $r\times r$ skew symmetric matrix $G$ \cite{Barouch1971,Cherng2006}.

Since the excitation probability
$p_k=\exp[-\pi(\gamma\sin k)^2/v]$
is invariant under the transformation $k \rightarrow \pi - k$, the correlator $\langle A_{\ell} B_{\ell+r} \rangle$ vanishes for odd $r$, reducing to $\delta_{r,0}$. Consequently, the spin correlation functions are nonzero only for even separations, $r = 2n$ with $n = 1, 2, \ldots$.  
As a result, the spin correlation functions can be expressed as products of sublattice correlators as follows:
%
\begin{equation}
\bl
\langle\sigma^{x}_{\ell}\sigma^{x}_{\ell+2n}\rangle 
=& \langle\langle\sigma^{x}_{\ell}\sigma^{x}_{\ell+n}\rangle\rangle\langle\langle\tau_{\ell}\tau_{\ell+n}\rangle\rangle,
\\
\langle\sigma^{y}_{\ell}\sigma^{y}_{\ell+2n}\rangle 
=& \langle\langle\sigma^{y}_{\ell}\sigma^{y}_{\ell+n}\rangle\rangle\langle\langle\tau_{\ell}\tau_{\ell+n}\rangle\rangle,
\el    
\end{equation}
%
where $\langle...\rangle$ and $\langle\langle...\rangle\rangle$ represent the expectation values on the full lattice and the sublattice, respectively~\cite{Cherng2006,Singh2021,Jafari2026a}.
Moreover, the sublattice correlators at separation $n$ can be written in terms of $n\times n$ determinants of Toeplitz matrices~\cite{Cherng2006,Singh2021}, 
%
\begin{equation}
\langle\langle\sigma^{x}_{\ell}\,\sigma^{x}_{\ell+n}\rangle\rangle = D_n[g^{+1,v}];
\quad
\quad
\langle\langle\sigma^{y}_{\ell}\,\sigma^{y}_{\ell+n}\rangle\rangle = D_n[g^{-1,v}].
\no
\end{equation}
%
Here, $g^{q,v}$ denotes the generating functions defined as
%
\begin{equation}
g^{q,v}(\lambda)=-(-\lambda)^q\,(1-2\,p_k),
\label{eq:gen_func0}
\end{equation}
%
where $\lambda=e^{2ik}$ and $D_n[g^{q,v}]$ are the corresponding Toeplitz matrix determinants for different sublattice spin correlators. 

Our primary interest lies in the behavior of the sublattice correlators at considerable separations, which relates to the large-$n$ asymptotics of Toeplitz determinants \cite{Barouch1971}. 
In the asymptotic limit, following Szeg\"{o}'s limit theorem, the Toeplitz determinant acquires the form~\cite{Cherng2006,Forrester2004,Marino2014,Singh2021,Jean2019,Abanov2003,Franchini2005,Dziarmaga2022,Ares2020,Albrecht1981},
%
\begin{equation}
D_n[g^{q,v} (\lambda)]\approx \exp\left[n\int^{ \pi}_0 \,\frac{d\theta}{ \pi}\, \mathrm{ln} \Big[
g^{q,v}(e^{i 2\theta})\Big]  \right].
\label{eq:Szego}
\end{equation}
%
%
Following Szeg\"{o}'s limit theorem, 
the sublattice correlators are given as 
$ e^{an}$,
where 
%
\begin{equation}
a= \int^{ \pi}_0  -(\lambda)^{q}\mathrm{ln}
[1-2p_k] \frac{dk}{\pi}.
\label{eq:Szego2}
\end{equation}
%
As expected, the transition to the upper level $p_k$ is depend on the value of $k$ and sweep velocity $v$. An adiabatic evolution
condition breaks down due to the crossing the critical (gap closing) point leading to a non-adiabatic transition with the maximum transition probability 
at gap closing modes $p_{\{k=0,\pi\}}=1$. Nevertheless, far away from the gap closing modes $k=0,\pi$, the system evolves adiabatically
due to the non-zero energy gap and leads to minimum transition probability at $k=\pi/2$.
Employing the Landau-Zener transition probability one finds that the inequality $p_{k=\pi/2}<1/2$ can be satisfied only for low sweep 
velocities, i.e., $v\leq v_{c}$ with $v_{c}=\pi \gamma^{2}/\ln[2]$~\cite{Sharma2016,Zamani2024,Jafari2024,Baghran2024}.
Therefore, for a sufficiently slow sweep $v\leq v_{c}$, there exist always two critical modes $k_{1}^{\ast}$ and $k_{2}^{\ast}=\pi-k_{1}^{\ast}$ where $p_{k_{1,2}^{\ast}} = 1/2$. 
It has been shown that the spin correlation functions at the end of the quench field undergo an abrupt transition at a critical sweep velocity, changing from {\it oscillatory behavior} for $v < v_c$ to {\it monotonic decay} for $v > v_c$, with a singularity at $v = v_c$ \cite{Cherng2006}.

The nonanalytic “singularity” (Fisher–Hartwig singularity \cite{Fisher1969,Kozlowska2019}) of the spin correlators  is tied to coalesce of two zeros of the Toeplitz generating 
function on the complex plane at $v_{c}$, not merely to the number of real momenta $k^{\ast}$ that satisfy ($p_{k^\ast}=1/2$). In other words, at $v=v_c$ a pair of complex singularities merges the unit circle, which produces the abrupt change in the Toeplitz asymptotics and hence the singular behaviour of the correlators. For ($v<v_c$) one indeed finds two real $k$-solutions ($p_k=1/2$), but those correspond to two separated (and generically simpler) singularities whose combined effect is an oscillatory contribution to correlations rather than the Fisher–Hartwig transition that causes the nonanalyticity at $v=v_c$. 
The apparently singular behavior is analogous to Stoke’s phenomenon \cite{Erdelyi1956,Meyer1989} for asymptotic series, where the coefficients
of an asymptotic expansion of a function may not be analytic in some parameters even when the function itself is
analytic in those parameters. 

%
%
\begin{figure*}
\begin{minipage}{\linewidth}
\centerline{\includegraphics[width=0.33\linewidth]{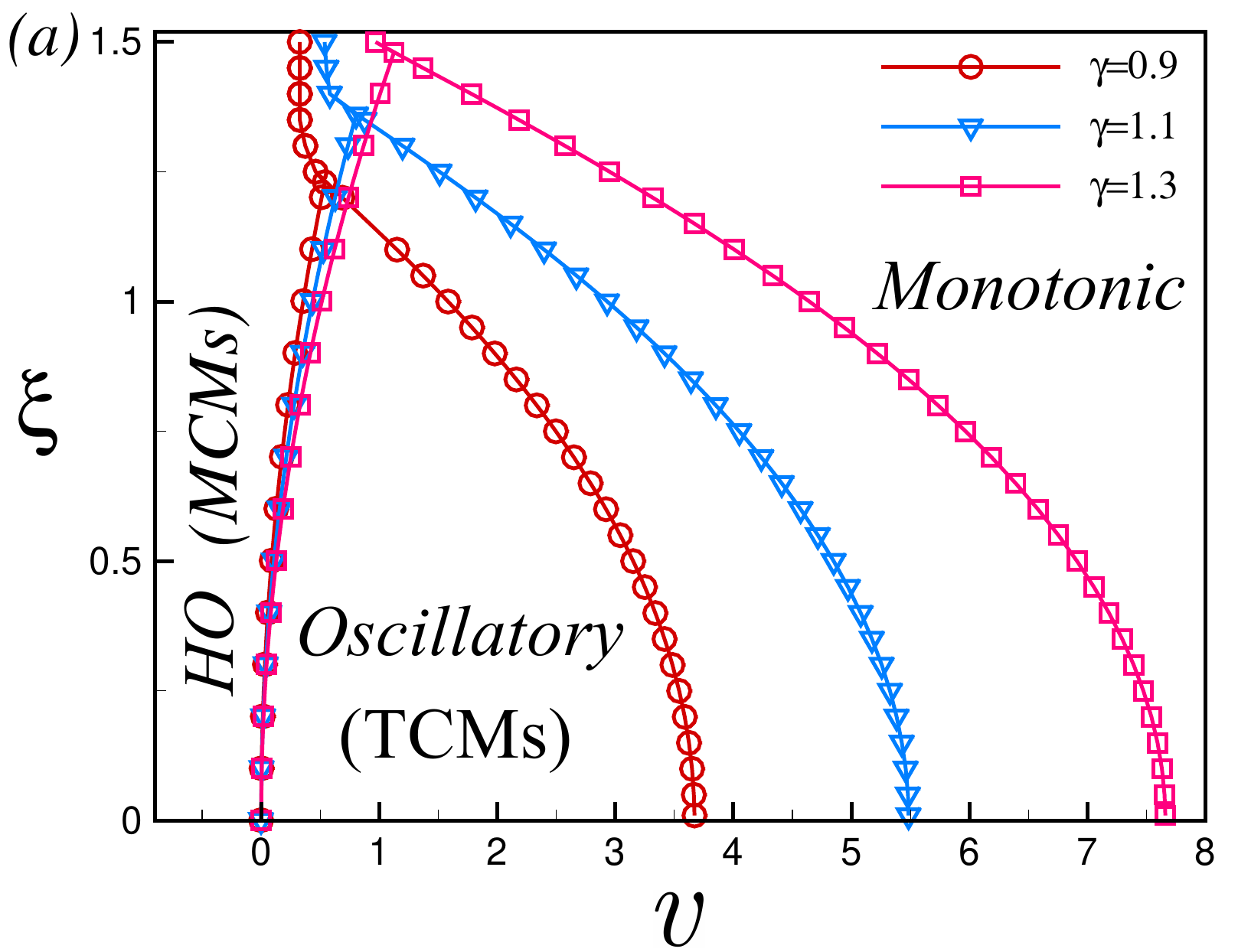}
\includegraphics[width=0.33\linewidth]{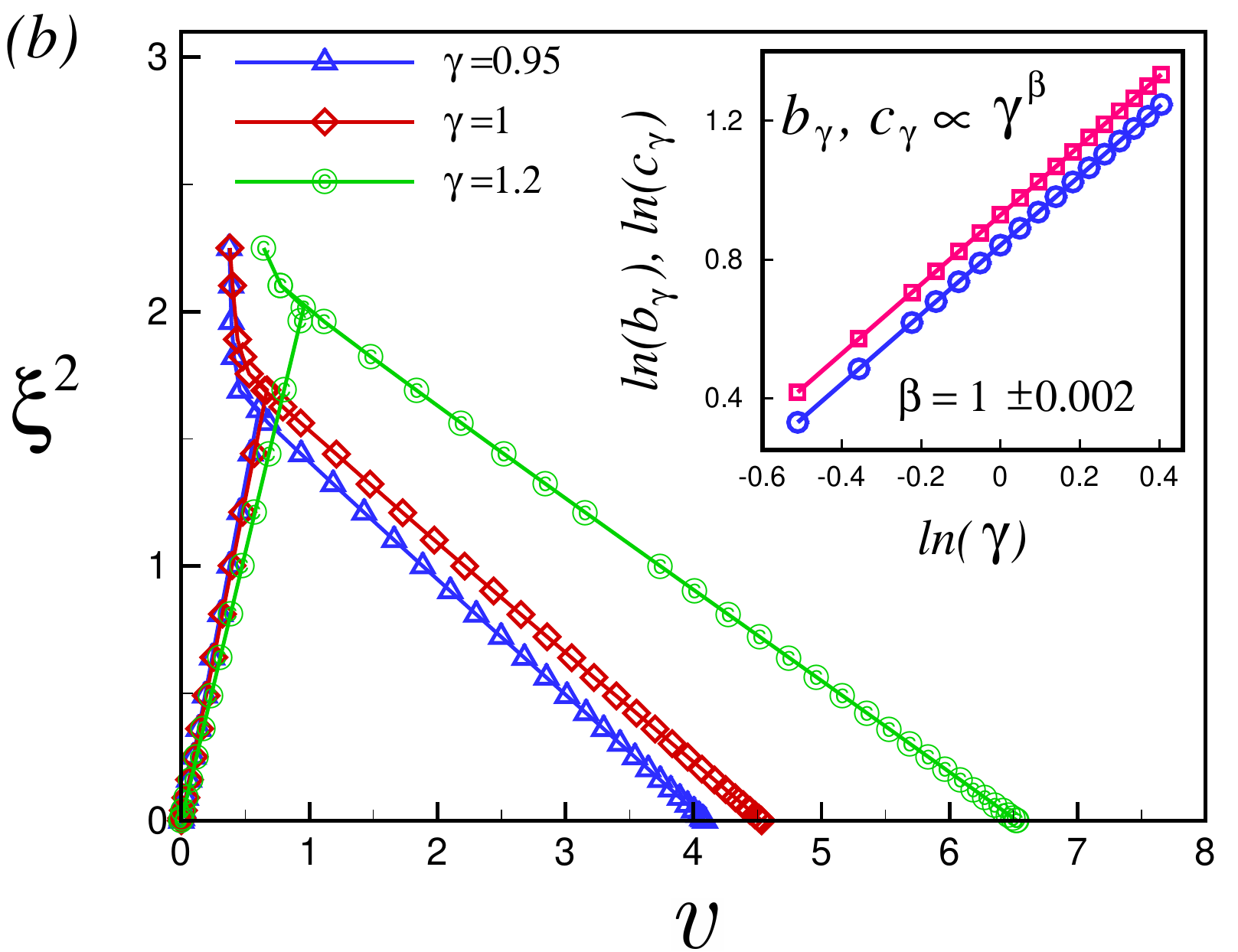}
\includegraphics[width=0.33\linewidth]{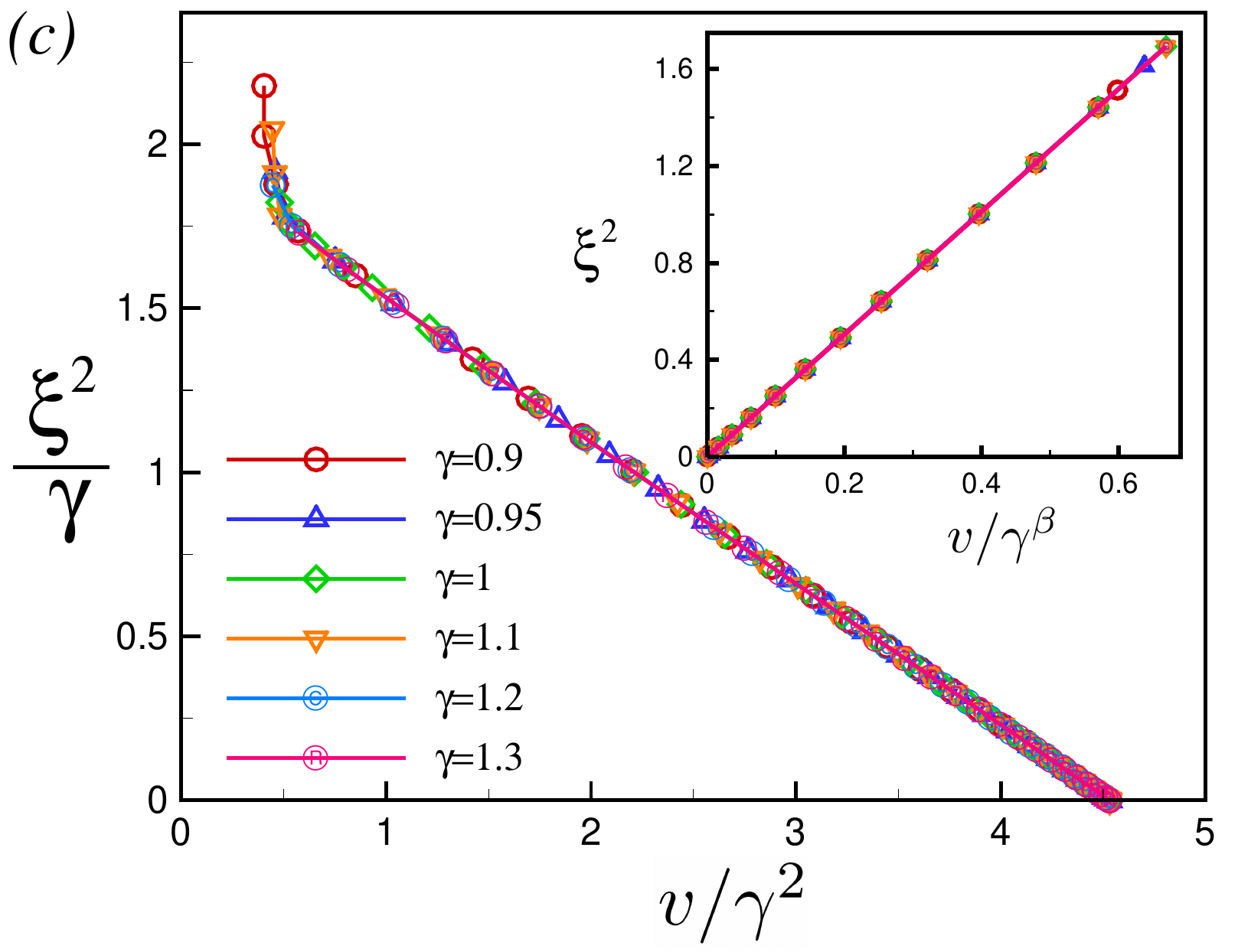}}
\centering
\end{minipage}
\vspace{-0.25cm}
\caption{
(a) Dynamical phase diagram of the $XY$ model in the $\xi$--$v$ plane for different values of the anisotropy $\gamma$, following a noisy quench across two quantum critical points at $h_c = -1$ and $h_c = 1$ ($h_i = -100$, $h_f = 100$). The critical sweep velocity $v_c$ separates regions in which spin correlation functions exhibit oscillatory and monotonic decay behavior. The oscillatory regime further splits into two distinct regions: a highly oscillatory region associated with multi-critical modes (MCMs) and an oscillatory region characterized by two critical modes (TCMs), distinguished by the multi-critical-mode velocity $v_m(\xi)$.
(b) The critical sweep velocity $v_c(\xi)$ and the multi-critical-mode velocity $v_m(\xi)$ are shown to scale linearly with the square of the noise strength, $\xi^2$. Inset: scaling of the slopes of $v_c(\xi)$ and $v_m(\xi)$, denoted by $b_\gamma$ and $c_\gamma$, respectively, as functions of the anisotropy $\gamma$.
(c) Scaling invariance of the critical sweep velocity in the presence of noise, $v_c(\xi)$, and of $v_m(\xi)$ (inset). Data for different values of the anisotropy collapse onto a single universal curve.
 }
\label{fig2}
\end{figure*}
%

{\em Noisy ramp quench.-}
We now examine whether the critical sweep velocity exhibits scaling and universality when noise is introduced into the driving field,
$h(t)=h_0(t)+R(t)$, where $R(t)$ denotes a random fluctuation confined to the ramp interval $[t_i,t_f)$ with vanishing mean,
$\langle R(t)\rangle=0$.
We consider Gaussian white noise characterized by the two-point correlation function
$\langle R(t)R(t')\rangle=\xi^2 \delta(t-t')$, where $\xi$ quantifies the noise strength.
Such white noise provides a good approximation to fast colored noise with exponentially decaying correlations, as described by the Ornstein--Uhlenbeck process \cite{Budini2000,Filho2017,Kiely2021,Dutta2016}. 
The numerical simulation indicates that, the summation $\sum_{k} e^{ikr}(\rho^{(d)}_{k,12})$ vanishes even when noise is present, 
$h(t) = vt+R(t) $, evolving from $h_i\ll h_c=-1$ to $h_f\gg h_c=1$.
As a result, Eq.~(\ref{eq:ABfunc2}) remains valid under noisy driving.

As in the noiseless case, the transition probabilities $p_k$ vary continuously with $k$ in the presence of noise; however, it is not a priori clear whether the specific value $p_k = 1/2$ is attained, or whether it may occur at multiple momenta. Near the gap-closing modes $k = 0,\pi$, the inequality $p_k^{\mathrm{max}} > 1/2$ is ensured by the Kibble--Zurek mechanism, which predicts a breakdown of adiabaticity as the system approaches criticality \cite{Kibble1976,Zurek1985,Bando2020,Mayo2021}.
On the other hand, noise generally enhances nonadiabatic transitions, raising the question of whether modes with $p_k^{\mathrm{min}} < 1/2$ persist. 
While closed-form expressions for transition probabilities are available in the absence of noise \cite{Vitanov1999,Zamani2024,Baghran2024}, no such analytical results exist when noise is present. 
This leads to the natural questions of whether noise can increase nonadiabatic transition probabilities sufficiently to violate the condition $p_k^{\mathrm{min}} < 1/2$, and whether the singular behavior of spin correlation functions survives or the critical sweep velocity is driven to zero.

To address these questions, we consider the exact master equation for the noise-averaged density matrix \cite{Budini2000,Filho2017,Kiely2021,Dutta2016},
\begin{equation}
\label{eq:master}
\frac{d}{dt}\rho_k(t)
=
-i\big[H^{(0)}_k(t),\rho_k(t)\big]
-\frac{\xi^2}{2}\big[H_1,\big[H_1,\rho_k(t)\big]\big],
\end{equation}
which governs the evolution of $\rho_k(t)$ associated with the Hamiltonian
$H_k(t)=H_{0,k}(t)+R(t)H_1$ during the quench interval $t\in[t_i,t_f)$, with $H_1=\sigma^z$.
By numerically solving this master equation, the mean transition probabilities are obtained as ensemble averages $\langle p_k\rangle$ over the noise realizations $\{R\}$.
In Fig.~\ref{fig1}(b)-(d) the transition probability has been plotted versus $k$ for different values of sweep velocity and the noise intensity
for a ramp from $h_i=-100$ to $h_f=100$.
As seen, when $\xi/v\ll1$, the main effect of noise is to shift the critical mode $k^{\ast}$.
By gradually increasing the noise intensity, the effect of noise takes a surprising turn for $\xi/v \sim \mathcal{O}(1)$. 
In this context, the critical momenta coalesce to a quasi-continuum, locking the $p_k$ curve to the value $1/2$ over a finite interval of momenta
leads to multi-critical modes (MCMs) with $p_k^{\ast}=1/2$. The locking of $p_k$ curve to $1/2$ over a finite range of momenta suggests that the noise acts like 
a high-temperature source, resulting in maximally mixed states unless the $k$-modes are too “light” (easily excited to the upper level by Kibble-Zurek Mechanism~\cite{Dutta2016}). It is worthwhile to mention that all critical modes $k^{\ast}$ in the multi-critical modes region are simple zeros and do not produce Fisher–Hartwig (square-root) singularities, and merely leading to a highly oscillatory behavior~\cite{Jafari2026a}.  
Physically, this behavior reflects an effective flattening of the excitation spectrum around $p_k = 1/2$, whereby noise generates a finite momentum window of maximally mixed modes.

Furthermore, the principal effect of noise is that the condition $p_{k=\pi/2} < 1/2$ can be satisfied only for sufficiently weak noise amplitudes. 
In the opposite regime $\xi/v \gg 1$, strong noise induces highly probable nonadiabatic transitions, preventing the emergence of a maximally mixed state ($p_k = 1/2$) at the end of the quench even for $v < v_c(\xi)$. 
As a result, the boundary separating oscillatory and monotonic regimes, where the spin correlation functions exhibit singular behavior, is shifted in the presence of noise.
The dynamical phase diagram of the model in the presence of noise is shown in Fig.~\ref{fig2}(a) in the $v$–$\xi$ plane for different values of the anisotropy $\gamma$.
As illustrated, the phase diagram consists of two primary regimes: oscillatory and monotonic.
Furthermore, the oscillatory regime is subdivided into two distinct regions: a noise-induced highly oscillatory region characterized by multi-critical modes for $v < v_m(\xi)$, and an oscillatory region with two critical modes for $v_m(\xi) < v < v_c(\xi)$.
The numerical results show that the critical sweep velocity $v_c(\xi)$, beyond which the oscillatory behavior crosses over to a monotonic decay, decreases with increasing noise intensity $\xi$.
These findings are consistent with the expectation that noise enhances nonadiabatic transitions, thereby suppressing the occurrence of a maximally mixed state ($p_k = 1/2$) at the end of the quench.
By contrast, the sweep velocity $v_m(\xi)$, below which the system enters the highly oscillatory regime, increases with increasing noise intensity.

For sufficiently strong noise ($\xi > 1$), the curve $v_m(\xi)$ intersects the critical sweep velocity $v_c(\xi)$, and the two merge into a single boundary.
Fig.~\ref{fig2}(b) shows the dynamical phase diagram in the $v$–$\xi^2$ plane.
As evident, for both weak ($\xi \sim \mathcal{O}(10^{-2})$) and strong ($\xi \sim \mathcal{O}(10^{-1})$) noise, the critical sweep velocity and $v_m(\xi)$ scale linearly with the square of the noise intensity, namely
$v_c(\xi) = - b_\gamma \xi^2 + v_c(0)$, where $v_c(0)$ denotes the critical sweep velocity in the absence of noise, and $v_m(\xi) = c_\gamma \xi^2$.

A more detailed analysis reveals that the slopes $b_\gamma$ and $c_\gamma$ exhibit power-law scaling with the anisotropy $\gamma$ [see the inset of Fig.~\ref{fig2}(b)], namely
$b_\gamma \propto \gamma^{\beta}$ and $c_\gamma \propto \gamma^{\beta}$, with an exponent $\beta = 1 \pm 0.002$.
These findings suggest a scaling form for $v_c(\xi)$ and $v_m(\xi)$, such that $v_c(\xi)$ and $v_m(\xi)$ are invariant under the transformations
$\xi \rightarrow \xi/\sqrt{\gamma}$, $v_c(\xi) \rightarrow v_c(\xi)/\gamma^{2}$, and $v_m(\xi) \rightarrow v_m(\xi)/\gamma^{\beta}$, respectively.
The scaling of the critical sweep velocity $v_c(\xi)$ and of $v_m(\xi)$ for different values of the anisotropy $\gamma$ is shown in Fig.~\ref{fig2}(c) and its inset, demonstrating that all curves collapse onto a single universal curve under the appropriate scaling transformation.
This data collapse provides clear evidence of the universality of spin correlation functions in the presence of noise.

{\em Summing up.-}
We have uncovered universal scaling behavior in nonequilibrium spin correlation functions under stochastic driving.
We showed that uncorrelated noise qualitatively reshapes the dynamical phase structure by reducing the critical sweep velocity at which spin correlations become singular, with a robust quadratic scaling in the noise strength.
Remarkably, when the noise intensity and sweep velocity are comparable, noise induces a finite momentum window of maximally mixed modes, giving rise to a highly oscillatory regime characterized by multi-critical dynamics.
Despite this strong modification of the excitation spectrum, the phase boundaries obey universal scaling laws and collapse onto a single curve under appropriate rescaling.
Our results demonstrate that noise can generate new universal dynamical structures rather than merely suppressing coherence,
thereby establishing noise as a constructive ingredient in nonequilibrium quantum critical phenomena.
Notably, the critical sweep velocity governing the singular behavior of spin correlations coincides with the velocity at which dynamical quantum phase transitions are wiped out \cite{Zamani2024,Jafari2024,Baghran2024,Ansari2025}.
\\

{\it Acknowledgements-}
A.A is supported by the Beijing Natural Science Foundation (Grant No. IS25015).


%


\begin{widetext}

\setcounter{figure}{0}
\setcounter{equation}{0}
\setcounter{section}{0}

\newpage

\section{Supplementary material}
\renewcommand\thefigure{S\arabic{figure}}
\renewcommand\theequation{S\arabic{equation}}
\setcounter{page}{1}

\vskip 0.5 cm

In this Supplemental material we elaborate on some technical aspects of the analysis presented in the main text \cite{Jafari2026bsm}, and also provide some background material.

\maketitle

\setcounter{figure}{0}
\setcounter{equation}{0}
\setcounter{section}{0}

\renewcommand\thefigure{S\arabic{figure}}
\renewcommand\theequation{S\arabic{equation}}

\subsection{Model and spin correlation functions}
We consider the one‐dimensional spin-$1/2$ XY model of length $N$ in a transverse magnetic field $h(t)$, described by the Hamiltonian  
%
\bea
\label{eqsm1}
{\cal H}_0(t) 
\!=\!
-\frac{J}{2} \sum_{j=1}^{N} \Big[\frac{1
\!+\!
\gamma}{2}\sigma_{j}^x \sigma_{j+1}^x \!+\!
\frac{1
\!-\!
\gamma}{2} \sigma_j^y \sigma_{j+1}^y
\!-\!
h_0(t) \sigma_j^z \Big],
\eea
%
where $\sigma^{\alpha=\{x,y,z\}}_{i}$ are the Pauli matrices acting on site $i$, and $\gamma$ is the anisotropy parameter.
We impose periodic boundary conditions.
In the case of a time-independent magnetic field ($h(t)=h$), the XY model exhibits an Ising‐type quantum phase transition at $\lvert h\rvert=1$,  and an anisotropic QPT along the line $\gamma=0$.
The Ising-like QPT occurs at $h=\pm1$ where the gap vanishes for $k=0$ and $k=\pi$ at the boundary between the paramagnetic phase and ferromagnetic phase.
By applying a Jordan–Wigner transformation \cite{LIEBsm,Jafari2012sm} followed by a Fourier transformation, the Hamiltonian in Eq.~\eqref{eqsm1} can be written as the sum of $N/2$ non-interacting terms 
%
\bea
\label{eqsm2}
{\cal H}(t) = \sum_{k} {\cal H}_{k}(t),
\eea
%
with
%
\bea
\label{eqsm3}
{\cal H}_{k}(t)
= &
\Big(h(t)-\cos(k)\Big)\Big(c_{k}^{\dagger} c^{}_{k}+c_{-k}^{\dagger} c^{}_{-k}\Big)-
{\it i}\gamma\sin(k)\Big(c_{k}^{\dagger} c_{-k}^{\dagger}+c^{}_{k} c^{}_{-k}\Big),
\eea
%
where $c_{k}^{\dagger}$ and  $c^{}_{k}$ are the spinless fermion creation and annihilation operators, respectively with $k= (2m-1)\pi/N$, for $m=1, 2, \dots, N/2$. 
Introducing the Nambu spinor $\mathds{C}^{\dagger}=(c_{k}^{\dagger},~c_{-k})$, the Hamiltonian ${\cal H}_{k}(t)$ can be expressed 
in Bogoliubov-de Gennes (BdG) form as
%
\bea
\label{eqsm4}
H_{k}(t)=\mathds{C}^{\dagger}{\cal H}_{k}(t)\mathds{C};
\quad
H_{k}(t)=\left(
\begin{array}{cc}
h_k(t) & -{\it i}\Delta_k \\
{\it i}\Delta_k & -h_k(t)\\
\end{array}
\right),
\eea
%
where $h_k(t)=h(t)-\cos(k)$ and $\Delta_k=\gamma\sin(k)$. Therefore the Bloch single particle Hamiltonian can be represented as
$H_{k}(t)=h_k(t)\sigma^{z}+\Delta_k\sigma^{y}$ with the corresponding eigenenergies $\varepsilon_k^{\pm}(t)=\pm\varepsilon_{k}(t)=\pm\sqrt{h^2_k(t)+\Delta^2_k}$.
In the fermion excitation formalism the instantaneous eigenvalues and eigenvectors of Hamiltonian Eq. (\ref{eqsm3}) in the even number of fermion sub-space ($|\alpha\rangle=|0\rangle$, $|\beta\rangle=c_{k}^{\dagger}c_{-k}^{\dagger}|0\rangle$) are expressed as
%
\bea
\bl
\label{eqsm5}
|\phi^{-}_{k}(t)\rangle
=&\cos(\frac{\theta_k(t)}{2})|0\rangle-i\sin(\frac{\theta_k(t)}{2})c_{k}^{\dagger}c_{-k}^{\dagger}|0\rangle
=\cos(\frac{\theta_k(t)}{2})|\alpha\rangle-i\sin(\frac{\theta_k(t)}{2})|\beta\rangle,
\\
|\phi^{+}_{k}(t)\rangle
=&-i\sin(\frac{\theta_k(t)}{2})|0\rangle+\cos(\frac{\theta_k(t)}{2})c_{k}^{\dagger}c_{-k}^{\dagger}|0\rangle
=-i\sin(\frac{\theta_k(t)}{2})|\alpha\rangle+\cos(\frac{\theta_k(t)}{2})|\beta\rangle
\el
\eea
%
where $\theta_k(t)=\arctan[\Delta_k/h_k(t)]$ and $|\alpha\rangle = (1\, 0)^T$, $|\beta\rangle = (0\, 1)^T$.
The two-point spin correlation functions can be expressed 
in terms of multipoint correlators of the JW fermions,
%
\bea
\bl
\label{eqsm6}
\langle\sigma^{x}_{\ell}\sigma^{x}_{\ell+r}\rangle =& \langle B_{\ell}A_{\ell+1}B_{\ell+1} \cdots A_{\ell+r-1}B_{\ell+r-1}A_{\ell+r}\rangle,
\\
\langle\sigma^{y}_{\ell}\sigma^{y}_{\ell+r}\rangle =& \langle A_{\ell}A_{\ell+1}B_{\ell+1} \cdots A_{\ell+r-1}B_{\ell+r-1}B_{\ell+r}\rangle,\\
\el
\eea
%
where
%
\begin{equation} 
\label{eqsm7}
A_j=c_j^\dag+c_j, \ \ B_j=c_j^\dag-c_j, \ \ j=1,\ldots , r+1.
\end{equation}
%
The expectation values in Eqs. (\ref{eqsm6}) are defined with respect to the time-evolved fermion states that stand in for the spin states after the JW transformation. 
For example,
%
\begin{equation} 
\label{eqsm8}
\langle A_1 B_2 \ldots \rangle = \mbox{Tr}(\rho(t) A_1 B_2 \ldots)
\end{equation}
%
with $\rho(t) = \otimes_k \rho_k(t)$, where $\rho_k(t)$ is the density operator for mode $k$ in a pure (mixed) ensemble of noise-free (noise-averaged) states from ramped quenches. 
Explicitly, let us consider the two-point function $\langle c_{\l}^\dagger c_{\l+r}^\dagger \rangle$, suppressing time arguments for
notational simplicity, and write
%
\begin{equation} 
\label{eqsm9}
\langle c_{\l}^\dagger c_{\l+r}^\dagger \rangle = \mbox{Tr}(\rho\, c_{\l}^\dagger c_{\l+r}^\dagger),
\end{equation}
%
By Fourier transforming, and considering $\rho(t) = \otimes_k \rho_k(t)$, where $\rho_k(t)$ is the density operator for mode $k$, one obtains from Eq. (\ref{eqsm9}),
%
\begin{eqnarray} 
\bl
\label{eqsm10}
\langle c^\dagger_{l} c^\dagger_{l+r}\rangle 
=& \frac{1}{N} \sum_{p,q} e^{-i(pl+q(l+r))} \langle  c^\dagger_p c^\dagger_q \rangle 
= \frac{1}{N} \sum_{p,q} e^{-i(pl+q(l+r))}\mbox{Tr}(\rho \,c^\dagger_p c^\dagger_q)   \\ 
=& \frac{1}{N} \sum_{k_j } e^{ik_jr}\,\mbox{Tr}((\otimes_{k \neq k_j} \rho_k)\otimes\rho_{k_j}c^\dagger_{k_j}c^\dagger_{-k_j})   
= \frac{1}{N} \sum_{k } e^{ikr}\,\mbox{Tr}(\rho_k\, c^\dagger_k c^\dagger_{-k}) 
\el
\end{eqnarray}
%
To take the trace in the last line of Eq. (\ref{eqsm10}) we use the diagonal bases $\{|\phi^\pm_k\}_k$ of the mode Hamiltonians at fixed time $t$,
%
\begin{equation} 
\label{eqsm11}
\mbox{Tr} (\rho_kc_{k}^\dagger c_{-k}^\dagger)=\langle\phi^{-}_k|\rho_kc_{k}^\dagger c_{-k}^\dagger|\phi^{-}_k\rangle + \langle\phi^{+}_k|\rho_kc_{k}^\dagger c_{-k}^\dagger|\phi^{+}_k\rangle.
\end{equation}

By inserting the unit operator $\mathbb{1} = |\phi^{-}_k\rangle\langle\phi^{-}_k| + |\phi^{+}_k\rangle\langle\phi^{+}_k|$ on the right-hand side of the equation, one obtains  
%
\begin{eqnarray} 
\label{eqsm12}
\mbox{Tr} (\rho_k c_{k}^\dagger c_{-k}^\dagger) &=& \langle\phi^{-}_k|\rho_k|\phi^{-}_k\rangle~\langle\phi^{-}_k|c_{k}^\dagger c_{-k}^\dagger|\phi^{-}_k\rangle
+ \langle\phi^{-}_k|\rho_k|\phi^{+}_k\rangle~\langle\phi^{+}_k|c_{k}^\dagger c_{-k}^\dagger|\phi^{-}_k\rangle \no \\
&+& \langle\phi^{+}_k|\rho_k|\phi^{-}_k\rangle~\langle\phi^{-}_k|c_{k}^\dagger c_{-k}^\dagger|\phi^{+}_k\rangle
+ \langle\phi^{+}_k|\rho_k|\phi^{+}_k\rangle~\langle\phi^{+}_k|c_{k}^\dagger c_{-k}^\dagger|\phi^{+}_k\rangle \no \\
& =& \rho^{(d)}_{k,11} ~\langle\phi^{-}_k|c_{k}^\dagger c_{-k}^\dagger|\phi^{-}_k\rangle + 
\rho^{(d)}_{k,12} ~\langle\phi^{+}_k|c_{k}^\dagger c_{-k}^\dagger|\phi^{-}_k\rangle\\
&+& \rho^{(d)}_{k,21}  ~\langle\phi^{-}_k|c_{k}^\dagger c_{-k}^\dagger|\phi^{+}_k\rangle 
+ \rho^{(d)}_{k,22}  ~\langle\phi^{+}_k|c_{k}^\dagger c_{-k}^\dagger|\phi^{+}_k\rangle, \no
\end{eqnarray}  
%
with the index $(d)$ a reminder that the density matrix is evaluated in the diagonal basis. The matrix elements of $c_{k}^\dagger c_{-k}^\dagger$ are easily calculated from Eqs. (\ref{eqsm5})-(\ref{eqsm6}), with the result
%
\begin{equation} 
\label{eqsm13}
\langle c_{\l}^\dagger c_{\l+r}^\dagger \rangle = \frac{1}{N}\sum_{k} e^{ikr}\Big[\cos^{2}(\frac{\theta_k(t)}{2})\rho^{(d)}_{k,12}+ \sin^{2}(\frac{\theta_k(t)}{2})\rho^{(d)}_{k,21} +\frac{i}{2}  (\rho^{(d)}_{k,11} - \rho^{(d)}_{k,22}) \Big].
\end{equation}
%
Similarly,
%
\begin{eqnarray} 
\label{eqsm14}
\langle c_{\l} c_{\l+r} \rangle &=& \frac{1}{N}\sum_{k} e^{ikr}\Big[\cos^{2}(\frac{\theta_k(t)}{2})\rho^{(d)}_{k,21}+ \sin^{2}(\frac{\theta_k(t)}{2})\rho^{(d)}_{k,12} -\frac{i}{2}  (\rho^{(d)}_{k,11} - \rho^{(d)}_{k,22}) \Big], \\
\label{eqsm15}
\langle c_{\l}^\dagger c_{\l+r} \rangle &=& \frac{1}{N}\sum_{k} e^{ikr}\Big[\cos^{2}(\frac{\theta_k(t)}{2})\rho^{(d)}_{k,22}+ \sin^{2}(\frac{\theta_k(t)}{2})\rho^{(d)}_{k,11} -\frac{i}{2}  (\rho^{(d)}_{k,12} - \rho^{(d)}_{k,21}) \Big], \\
\label{eqsm16}
\langle c_{\l} c_{\l+r}^\dagger \rangle &=& \frac{1}{N}\sum_{k} e^{ikr}\Big[\cos^{2}(\frac{\theta_k(t)}{2})\rho^{(d)}_{k,11}+ \sin^{2}(\frac{\theta_k(t)}{2})\rho^{(d)}_{k,22} -\frac{i}{2}  (\rho^{(d)}_{k,21} - \rho^{(d)}_{k,12}) \Big].
\end{eqnarray}
%
By feeding the expressions from Eqs. (\ref{eqsm13})-(\ref{eqsm16}), we finally obtain the $A$ and $B$ two-point functions that determine the spin correlations in Eq. (\ref{eqsm7}). Pruning the outcome, the result is
 
%
\begin{eqnarray} 
\no
\langle A_{\ell} A_{\ell+r} \rangle &=& \langle c_l^\dag c_{\l+r}^\dag \rangle + \langle c_l c_{\l+r} \rangle + \langle c_l^\dag c_{\l+r} \rangle + \langle c_l c_{\l+r}^\dag \rangle =
\frac{1}{N} \sum_{k} e^{ikr} \mbox{Re}(\rho^{(d)}_{k,12}(t)) + \delta_{r,0}, \\
\label{eqsm17}
\langle B_{\ell} B_{\ell+r} \rangle &=& \langle c_l^\dag c_{\l+r}^\dag \rangle + \langle c_l c_{\l+r}\rangle - \langle c_l^\dag c_{\l+r} \rangle - \langle c_l c_{\l+r}^\dag \rangle = \frac{1}{N} \sum_{k} e^{ikr} \mbox{Re}(\rho^{(d)}_{k,12}(t)) - \delta_{r,0}, \\
\no
\langle A_{\ell} B_{\ell+r} \rangle &=& - \langle B_{\ell+r} A_{\ell} \rangle = \langle c_l^\dag c_{\l+r}^\dag \rangle - \langle {c_l}{c_{\l+r}} \rangle - \langle c_l^\dag c_{\l+r} \rangle + \langle c_l c_{\l+r} ^\dag  \rangle
\\
\no
&=& \frac{1}{N} \sum_{k} e^{ikr} e^{i\theta_k(t)}\Big[2 i \mbox{Im}(\rho^{(d)}_{k,12}(t))+ \rho^{(d)}_{k,11}(t)-\rho^{(d)}_{k,22}(t) \Big],
\end{eqnarray}
%
Let us mention that the two-point fermion functions above reduce to those of the equilibrium case if we take $\rho^{(d)}_{k,22}=\rho^{(d)}_{k,12}=\rho^{(d)}_{k,21}=0$, and 
$\rho^{(d)}_{k,11}=1$:
%
\bea
\no
\langle A_{l}A_{l+r} \rangle = \delta_{r,0}, \ \ \ \langle B_{l}B_{l+r} \rangle = -\delta_{r,0}, \ \ \
\langle A_{l} B_{l+r} \rangle = -\langle B_{l+r} A_{l} \rangle = \frac{1}{N}\sum_{k}e^{ikr} e^{i\theta_k(t)}.
\eea
%
In the thermodynamic limit $N\rightarrow\infty$, this can be written as 
%
\bea
\no
\langle {{A_l}{A_{l+r}}} \rangle = \delta(r), \ \ \ \langle {{B_l}{B_{l+r}}} \rangle  -\delta(r), \ \ \
\langle A_{l} B_{l+r} \rangle = -\langle B_{l+r} A_{l} \rangle = \frac{1}{4\pi}\int_{-\pi}^{\pi} e^{ikr} e^{i\theta_k(t)}dk,
\eea
%
in agreement with Ref. \cite{Franchini2017}. 

Also note that for a quench from $h(t_{\text{initial}})\!\rightarrow\! -\infty$ to $h(t_{\text{final}})\!\rightarrow \!+\infty$, the off-diagonal terms $\rho^{(d)}_{k,12}(t)=(\rho^{(d)}_{k,21})^{\ast}$ of the density matrix $\rho_{k}(t)$ vanish, as does $\theta_{k}(t)$. 
As a consequence, the two-point fermion functions in Eq. (\ref{eqsm17}) reduce to 
%
\bea
\no
\langle {{A_l}{A_{l+r}}} \rangle = \delta_{r,0},~~
\langle {{B_l}{B_{l+r}}} \rangle = -\delta_{r,0}, ~~
\langle A_{l} B_{l+r} \rangle = -\langle B_{l+r} A_{l} \rangle = \frac{1}{N}\sum_{k}  e^{ikr} (1-2\rho^{(d)}_{k,22}),
\eea
%
which in the large-time thermodynamic limit reproduces the result in Refs. \cite{Cherng2006sm,Sengupta2009sm,Nag2011sm}, 
%
\bea
\no
\langle {{A_l}{A_{l+r}}} \rangle &=& \delta(r), \ \ \ 
\langle {{B_l}{B_{l+r}}} \rangle = -\delta(r), \\
\langle A_{l} B_{l+r} \rangle&=&-\langle B_{l+r} A_{l} \rangle = \frac{1}{2\pi} \int_{-\pi}^{\pi} e^{ikr} (1-2p_{k}(t))= 
-\frac{2}{\pi}\int_{0}^{\pi}p_{k}(t)\cos(kr)dk,
\eea
%
where $\mbox{lim}_{t\rightarrow\infty}\, \rho^{(d)}_{k,22}(t) = p_k=\exp(-\pi(\gamma\sin(k))^2/v)$.

\subsection{Spin correlators and Szeg\"o's limit theorem}

As shown in Ref.~\cite{Cherng2006sm}, the spin correlation functions can be expressed as products of
sublattice correlators. The sublattice correlators at separation $n$ can, in turn, be written as
Toeplitz determinants,
\begin{equation}
\langle\langle \sigma^{x}_{\ell}\sigma^{x}_{\ell+n} \rangle\rangle
=
D_n\!\left[g^{+1,v}\right],
\end{equation}
\begin{equation}
\langle\langle \sigma^{y}_{\ell}\sigma^{y}_{\ell+n} \rangle\rangle
=
D_n\!\left[g^{-1,v}\right],
\end{equation}
where $g^{q,v}$ with $q=\pm1$ denotes the generating function, defined as
\begin{equation}
g^{q,v}(\lambda)
=
-(-\lambda)^q\,(1-2p_k),
\label{eq:gen_func0}
\end{equation}
with $\lambda=e^{2ik}$.
The quantities $D_n[g^{q,v}]$ are the corresponding Toeplitz determinants associated with the
different sublattice spin correlators.

Given a generating function $g^{q,v}(\lambda)$, the Toeplitz determinant $D_n[g^{q,v}]$ is defined
as~\cite{Cherng2006sm,Marino2014sm}
\begin{equation}
D_n[g^{q,v}]
=
\begin{vmatrix}
f^{(q)}_0      & f^{(q)}_{-1}    & \cdots & f^{(q)}_{-(n-1)} \\
f^{(q)}_1      & f^{(q)}_{0}     & \cdots & f^{(q)}_{-(n-2)} \\
\vdots         & \vdots          & \ddots & \vdots          \\
f^{(q)}_{n-1}  & f^{(q)}_{n-2}   & \cdots & f^{(q)}_{0}
\end{vmatrix},
\label{eq:T_determinant}
\end{equation}
where the matrix elements $f^{(q)}_l$ are the Fourier coefficients of the generating function,
defined through the expansion
\begin{equation}
g^{q,v}(\lambda)=\sum_l f^{(q)}_l\,\lambda^l .
\end{equation}
These coefficients are obtained via the contour integral
\begin{equation}
f^{(q)}_l
=
\oint_C \frac{d\lambda}{2\pi i\,\lambda}\,
\lambda^{-l}\, g^{q,v}(\lambda),
\label{eq:gen_func1}
\end{equation}
where $C$ denotes the unit circle $|\lambda|=1$.

Expressed in terms of the momentum variable $k$, this integral takes the form
\begin{equation}
f^{(q)}_l
=
\int_{-\pi}^{\pi} \frac{dk}{\pi}\,
e^{-i2kl}\, g^{q,v}(e^{i2k}) .
\label{eq:gen_z_T_elements}
\end{equation}

\subsection{Simple zeros and Fisher--Hartwig singularities of the generating functions}

In the asymptotic large-separation limit, Szeg\"o's theorem implies that the Toeplitz determinant takes the form~\cite{Cherng2006,Forrester2004,Marino2014,Singh2021,Jean2019,Abanov2003,Franchini2005,Dziarmaga2022,Ares2020,Albrecht1981}
\begin{equation}
D_n\!\left[g^{q,v}(\lambda)\right]
\sim
n^{\sum_z (\alpha_z^2-\beta_z^2)}
\exp\!\left[
n\!\int_0^{\pi}\!\frac{d\theta}{\pi}\,
\ln\!\big(\lambda\, g^{q,v}(\lambda)\big)
\right],
\label{eq:Szego}
\end{equation}
where $g^{q,v}(\lambda)$ is the generating function defined in Eq.~(\ref{eq:gen_func0}), and
$\alpha_z$ and $\beta_z$ are the Fisher--Hartwig exponents associated with the zeros and phase
discontinuities of $g^{q,v}(\lambda)$~\cite{Fisher1969sm,Basor1991sm,Ehrhardt2001sm,Deift2011sm,Deift2013sm}.

The exponent $\alpha_z$ characterizes the order of a zero of the generating function,
\begin{equation}
g^{q,v}(\lambda)
=
\mathrm{const.}\,|k-k^\ast|^{\alpha_z/2}
+
\mathcal{O}\!\left(|k-k^\ast|^{\alpha_z/2+1}\right),
\end{equation}
where $k^\ast$ denotes a critical momentum at which $g^{q,v}(\lambda)$ vanishes.
The exponent $\beta_z$ measures a phase discontinuity of the generating function: if
$g^{q,v}(\lambda)$ crosses zero without a phase jump, $\beta_z=0$, whereas a sign change
corresponds to $\beta_z=\pm 1/2$.

In the absence of noise, the excitation probability is given by the Landau--Zener formula
$$p_k=\exp[-\pi(\gamma\sin k)^2/v].$$
For sweep velocities $v<v_c$~\cite{Jafari2026bsm}, there always exist two critical modes
$k_1^\ast$ and $k_2^\ast=\pi-k_1^\ast$ satisfying $p_{k_{1,2}^\ast}=1/2$, at which the
generating function vanishes.
A straightforward expansion around these modes yields
$$
g^{q,v}(k)\propto |k-k^\ast|,
$$
implying $dp_k/dk|_{k^\ast}\neq0$ and hence $\alpha_z=1/2$.
Furthermore, the phase of the generating function remains continuous at these points,
leading to $\beta_z=0$.
According to the Fisher--Hartwig theorem, such simple zeros generate oscillatory exponential
decay of the spin correlation functions~\cite{Fisher1969sm,Basor1991sm,Ehrhardt2001sm,Deift2011sm,Deift2013sm}.

By contrast, at the critical sweep velocity $v=v_c$, the two critical modes merge at
$k^\ast=\pi/2$.
In this case, the generating function behaves as
$g^{q,v}(k)\propto |k-k^\ast|^2$, corresponding to $dp_k/dk|_{k^\ast}=0$ and hence
$\alpha_z=1$, while the phase remains continuous so that $\beta_z=0$.
This higher-order zero produces a Fisher--Hartwig singularity, leading to a divergence of
the asymptotic spin correlation length~\cite{Fisher1969sm,Basor1991sm,Ehrhardt2001sm,Deift2011sm,Deift2013sm}.

In the presence of noise, the transition probabilities and the corresponding critical modes
are obtained numerically.
The exponents $\alpha_z$ are extracted from the scaling behavior of the generating function
in the vicinity of the critical modes, while $\beta_z$ is determined from its phase
structure around these points.

%


\end{widetext}

\end{document}